\documentclass[12pt]{article}
 \usepackage[dvips]{graphicx}
 \usepackage[dvips]{graphics}
 \usepackage{amsmath}
 \usepackage{enumerate}
 \usepackage{psfrag}

 \newcommand{\newc}{\newcommand}
 \newc{\ra}{\rightarrow}
 \newc{\lra}{\leftrightarrow}
 \newc{\beq}{\begin{equation}}
 \newc{\eeq}{\end{equation}}

\begin{document}

 \begin{center}
 {\LARGE \bf  Tokamak MHD equilibria  with \\
 reversed magnetic shear and sheared flow}
 \footnote{A preliminary version of the present study was presented
 in the 29$^{th}$ EPS Conference on Plasma Phys. and Control.
 Fusion, Montreux, 17-21 June 2002 \cite{PoTh}}

 \vspace{2mm}

 {\large G. Poulipoulis$^\dag$\footnote{me00584@cc.uoi.gr}, G. N. Throumoulopoulos$^\dag$\footnote{gthroum@cc.uoi.gr}, H.
 Tasso$^\star$\footnote{het@ipp.mpg.de}}

 \vspace*{1mm}

 $^\dag${\it University of Ioannina, Association Euratom - Hellenic
 Republic,\\ \vspace{-1mm}
 Section of Theoretical Physics, GR 451 10 Ioannina, Greece}

 \vspace{1mm} \noindent $^\star${\it  Max-Planck-Institut f\"{u}r
 Plasmaphysik, Euratom Association,\\ \vspace{-1mm}
 D-85748 Garching, Germany }
 \end{center}

 \begin{center}
 {\bf Abstract}
 \end{center}

 Analytic solutions of the magnetohydrodynamic equilibrium
 equations for a cylindrically symmetric magnetically confined
 plasma with reversed magnetic shear, $s<0$,  and sheared  flow  are
 constructed by prescribing the safety factor-, poloidal velocity-
 and axial velocity- profiles consistently with experimental
 ones. On the basis of the solutions obtained in most of the cases considered
 it turns out that an
 increase of $|s|$ and of the velocity components  result in larger absolute values for the radial electric field,
 $E_r$, its shear,  $|dE_r/dr|\equiv |E_r^\prime|$,  and the  ${\bf E}\times {\bf B}$
 velocity shear, $\omega_{{\bf E}\times {\bf  B}}=|d/dr({\bf E}\times{\bf B}/B^2)|$, which may play a role
 in the formation of Internal
 Transport Barriers (ITBs) in tokamaks. In particular for a constant axial magnetic field,
 $\omega_{{\bf E}\times{\bf B}}$ at the point where $E_r^\prime=0$ is proportional to $1-s$. Also,
 $|E_r^\prime|$  and $\omega_{{\bf E}\times{\bf B}}$ increase as the velocity shear takes larger values.
 The results clearly
 indicate that $s<0$ and  sheared flow act synergetically in the
 formation of ITBs with the impact of the flow, in particular the
 poloidal one, being stronger than that of $s<0$.
\newpage

\begin{center}
 {\large \bf 1. \  Introduction}
\end{center}

 Understanding Internal Transport Barriers (ITBs) in plasmas is very important
 for the advanced tokamak scenarios \cite{HaPr},\cite{sak}. The ITBs  usually are associated
 with reversed magnetic shear profiles \cite{str},\cite{con} and their main characteristics
 are  steep pressure profiles in the barrier region \cite{lev} and radial
 electric fields associated with sheared flows \cite{tal},  \cite{CaWa}. The
 mechanism responsible for the formation of ITBs is far for completely
 understood. It is believed that the flow, the radial electric field,
 its shear
and the ${\bf E}\times {\bf B}$ velocity shear,
 \beq
 \omega_{{\bf E}\times {\bf B}}=\left
 |\frac{d}{dr} \frac{{\bf E}\times {\bf B}}{B^2}\right |,
 \label{5}
 \eeq
 play a role in the barrier formation by mode
 decorrelation thus resulting in a reduction of the outward particle
 and energy transport \cite{sak}, \cite{bur}, \cite{Te}.

 The experimental evidence up to date  has not made clear  whether the reversed magnetic
 shear, $s<0$, or the sheared flow (toroidal or poloidal) are more important for the ITBs formation.
 In some experiments the safety factor profile is considered as the crucial quantity (e.g. \cite{eri})
 while according to  others the necessity of
 reversed magnetic shear is questionable (e.g.  \cite{bur}). On the other hand,
the flow-either toroidal \cite{cri} or poloidal
\cite{rbel},\cite{ric} may be important in the formation of
 ITBs. Also,  it has been argued that the toroidal velocity may be more
 important than the poloidal one (see for example Ref. \cite{rbel}). It should be noted, however,
 that only few direct measurements of
 the poloidal velocity have been performed; this velocity is
 usually calculated  by means  of  neoclassical theory \cite{cri}.

 The aim of the present  work is to contribute to the answer of the above mentioned open questions by
 studying magnetohydrodynamic (MHD) cylindrical equilibria
 with reversed magnetic shear and sheared flow. The study can be viewed as an extension of a previous one
 on tokamak equilibria with incompressible sheared flows and monotonically
 increasing q-profiles  in connection with certain characteristics of the L-H
 transition \cite{sim}. The work is conducted through the following steps: The profiles of certain free quantities,
 including the safety factor and the velocity components are first
 prescribed and then exact equilibrium solutions are  constructed
 self consistently. This is the subject of Sec. 2. In Sec. 3
 on the basis of the solutions obtained  the equilibrium properties  are examined and the
 impact of $s<0$ and the flow on  $E_r$, $E_r^\prime$ and $\omega_{{\bf E}\times{\bf B}}$ is evaluated.
 The conclusions are
 summarized in Sec. 4.

 \begin{center}
{\large \bf 2. \  Cylindrical equilibria with reversed magnetic shear}
 \end{center}

 The equilibrium of a cylindrical plasma with flow satisfies (in
 convenient units) the relation
 \beq
 \frac{d}{dr}\left(P + \frac{B_\theta^2+ B_z^2}{2}\right)+
 \left(1-M_\theta^2\right)\frac{B_\theta^2}{r}=0
 \label{1}
 \eeq
 stemming from the radial component of the force-balance equation
 $\rho ({\bf v} \cdot {\bf\nabla})  {\bf v} = {\bf j} \times {\bf
 B} - {\bf\nabla} P $ with the aid of Amp$\acute{e}$re's law. Here, $P$ is the plasma  pressure;
 $B_\theta$ and $B_z$ are the poloidal and axial components of the
 magnetic field, respectively; $M_\theta^2=(v_\theta^2
 \varrho)/B_\theta^2$ is the square of the Mach number defined as
 the ratio of the poloidal velocity to the poloidal-magnetic-field
 Alfv\'en velocity. Because of the symmetry any equilibrium
 quantity depends only on the radial distance $r$ and the axial
 velocity $v_z$ as well as the velocity shear do not appear in
 (\ref{1}); also, the flow is incompressible. In addition to $v_z$
 four out of the five quantities in (\ref{1}) can be prescribed.

\begin{figure}[!h]
 \begin{center}
 \psfrag{q}{$q(\rho)$}
 \psfrag{r}{$\rho$}
 \psfrag{W}{ WRS }$-$
 \psfrag{S}{SRS $\cdots$}
 \includegraphics[scale=1]{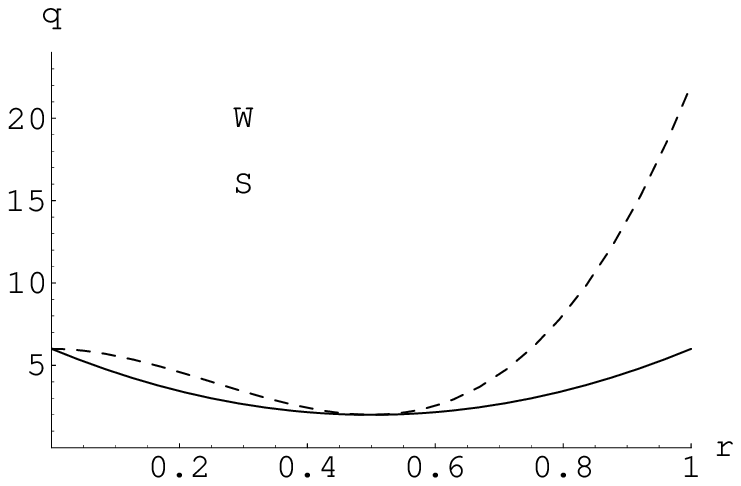}
 \caption{}
 \label{fig:1}
 \end{center}
 \end{figure}

 On account of typical experimental ITB profiles we prescribed the quantities $q$, $B_z$,
 $v_\theta$, $v_z$ and $\varrho$ as follows: \\
  strongly reversed shear profile (SRS)  (Fig. \ref{fig:1})
 \beq
 q(\rho)=q_c\left(1-\frac{3\Delta q}{q_c}\frac{r_0^2}{r_{\min}^2}\rho^2
 +\frac{2\Delta q}{q_c} \frac{r_0^3}{r_{\min}^3} \rho^3 \right)
 \label{2}
 \eeq
 or alternatively weakly  reversed shear  profile (WRS)
 \beq
 q(\rho)=q_c\left(1-\frac{2\Delta q}{q_c}\frac{r_0}{r_{\min}}\rho
 +\frac{\Delta q}{q_c} \frac{r_0^2}{r_{\min}^2} \rho^2\right)
 \label{3}
 \eeq
 where $\rho=r/R_0$ with $r_0$ defining the plasma surface,
 $q_c=q(r=0)$, $r_{\min} $ is the position of minimum $q$, and
 $\Delta q=q_c-q_{\min}$. The SRS-profile (\ref{2}) does exhibit a
 maximum at the plasma center $r=0$ in addition to the minimum one at
 $r=r_{\min}$ and has stronger magnetic shear in the central region
 just inside the  $q_{\min}$ position than that of the WRS one.  It should be clarified, however, that
 the WRS profile (\ref{3}), which does not  have an extremum on the magnetic axis
$r=0$, has been chosen  in order to simplify the calculations
though the physical situation may not be well represented in the
immediate vicinity of the magnetic axis; \beq
 B_z= B_{z0} \left\lbrack 1+ \delta (1-\rho^2)\right \rbrack ^{1/2}
 \label{31}
 \eeq
 where $B_{z0}$ is the vacuum magnetic field and the parameter
 $\delta$ is related to the magnetic properties of the plasma, i.e.
 for $\delta<0$ the plasma is diamagnetic;\\
 Gaussian-like poloidal velocity profile
 \beq
 v_\theta= 4 v_{\theta 0}\rho(1-\rho)\exp \left\lbrack -\frac{(\rho -\rho_{min})^2}{h}
 \right\rbrack
 \label{4a}
 \eeq
 where the parameter $h$ determines its broadness and
 $v_{\theta 0}$ is the maximum of $v_\theta$;
 either peaked axial velocity profile
 \beq
 v_z= v_{z0} (1-\rho^3)^3
 \label{4b}
 \eeq
 or Gaussian-like $v_z$ profile similar to that of (\ref{4a});
 and the  density profile
 \beq
 \varrho = \varrho_0(1-\rho^3)^3.
 \label{3a}
 \eeq

 %

 The following quantities can then be calculated: the poloidal
 magnetic field $B_\theta = {\cal \epsilon}\rho B_z/ q$ where
 ${\cal \epsilon} =r_0/R_0$ is the inverse aspect ratio with $2\pi
 R_0$ associated with  the length of the plasma column; the
 magnetic shear $s= (r/q)(dq/dr)$; the current density via Ampere'
 s law; the electric field via Ohm' s law; its shear $E_r^\prime$
 and $\omega_{{\bf E}\times{\bf B}}$ by (\ref{5}). Also,
 integration of (\ref{1}) so that $P(r=r_0)=0$ yields the pressure.
  The calculations have been performed analytically by developing a
 programme for symbolic computations \cite{Poul} in connection with Ref.
 \cite{Math}. This also allowed us to examine conveniently
 purely poloidal flows, purely axial  flows, $z$-pinch configurations or
 $\theta$-pinch configurations as particular cases.  The analytic expressions  which can be derived
 readily by the programme are generally lengthy and will not
 be given explicitly here. Some concise and instructive
 expressions will only be presented in the next section along with
 typical profiles for the calculated quantities supporting the results
 obtained.

 \begin{center} {\large \bf 3.\  Results}
 \end{center}

 We have set the following values for some of the parameters:
 $B_{z0}= 1$ Tesla, $\varrho_0= 8.35 \times 10^{-8} \mbox{kgr/m}^3$ corresponding to
 $n_0=5\times 10^{19}$  particles/m$^3$, $\rho_{\min}=0.5$, $ {\cal \epsilon}=r_0/R_0\approx 1/3$,
 $\delta=-0.0975$,  $q_{\min}=2$,
$\max v_{\theta}= 1\times 10^4$ m/sec and  $\max v_{z}= 1\times
10^5$
 m/sec; Consequently, it is guaranteed that $M_\theta^2
 \approx M_z^2$, where $M_z^2= (v_z^2 \varrho )/ B_z^2$, a scaling
 typical in tokamaks because $B_z\approx 10\ B_\theta$ and
 $v_z\approx 10\ v_\theta$. It is  noted here that since in
 tokamaks $M_\theta<0.1$ the flow term in (\ref{1}) is perturbative
 around the ``static" equilibrium $M_\theta=0$. Also, the choice  $q_{\min}=2 $
was made because
according to experimental evidence for $q_{\min}<2$ strong MHD activity
 destroys confinement   possibly due to a  double tearing mode \cite{wol}.
 A similar result was found numerically for one-dimensional cylindrical equilibria with hollow currents
 in Ref. \cite{KeTa}. The impact of the magnetic shear and flow on
 the equilibrium, in particular
 on the quantities  $E_r$, $E_r^\prime$ and $\omega_{{\bf E}\times{\bf B}}$, was examined  by varying the
 parameters $q_c$, $\Delta q$, $h$, $v_{z0}$, and $v_{\theta 0}$ [Eqs.
 (\ref{2}),
 (\ref{3}), (\ref{4a}) and (\ref{4b})].

 For reversed magnetic shear profiles we came to the following
 conclusions:\\

 \noindent
 1. {\large \em Pressure }\\

 Substitution of $B_\theta$ and its derivative in terms of
 $q$ and $s$ in  (\ref{1}) yields
 \beq
 P^\prime= -B_zB_z^\prime \left\lbrack 1+ \left({\cal \epsilon}
 \frac{r_0}{R_0}\right)^2\right\rbrack +r_0\rho\left\lbrack M_\theta^2 +  (s-2)\right\rbrack
 \left(\frac{B_z}{R_0q}\right)^2.
 \label{6}
 \eeq
 For $s<0$, increase of $|s|$ makes the pressure profile steeper
 (see also Fig. 2).
 Equation (\ref{6}) also implies that the pressure profile becomes steeper when the plasma is
 more diamagnetic, i.e. when $B_z^\prime$ related to the parameter $\delta$ in (\ref{31})
 takes larger values.
 \vspace{3mm}

 \noindent
 2. {\large \em Current density}

 \begin{itemize}
 \item  The axial current density profile  becomes hollow and,  irrespective of the reversal of the magnetic shear,
 there is a critical distance $\rho_{cr}$ outside  the $q_{\min}$
 position at which $J_z$ becomes negative (Fig. 3).
 In particular, for $B_z=B_{z0}= \mbox{const.}$ one obtains
 \beq
 J_z= \frac{1}{r}\frac{d}{dr}(rB_\theta)=\frac{B_{z0}}{R_0q}(2-s)
 \label{7}
 \eeq
 Consequently, for $s>2$,  $J_z$ reverses.
 The radial distances at which $J_z=0$ for the  SRS [Eq. (\ref{2})]
 and the WRS [Eq. (\ref{3})] q-profiles, respectively, are
 $$
 \rho_{cr}^{SRS}=\rho_{\min}\left(\frac{q_c}{\Delta
 q}\right)^{1/3}
 $$
 and
 $$
 \rho_{cr}^{WRS}=\rho_{\min}\frac{q_c}{\Delta q}.
 $$
 Therefore, the position of $\rho_{cr}$ is shifted towards the center as $s$
 takes lower negative values. It is noted here that equilibrium
 toroidal current density reversal for monotonically increasing
 $q$-profiles was reported in Ref. \cite{Ma} (Fig. 3  therein).

 \item Very large values of $\Delta q $ on the order of $10^2$ result in the formation
 of $j_z$ profiles with "holes" in the central region- $j_z\approx
 0$- inside the $\rho_{\min}$ position  as demonstrated in Fig. 4,
 a result consistent with experimental evidence (\cite{fuj},
 \cite{haw}).
 \item The total axial current $I_z=2\pi r_0 B_\theta(r_0)$ for SRS
 profiles is smaller than that for WRS profiles.
 \end{itemize}

 \noindent
 3. {\large \em $E_r$ and $E_r^\prime$}
 \begin{itemize}
 \item Typical $E_r$ profiles exhibit an extremum
 in the region around $q_{min}$ and vanish at $\rho =0$ and $\rho =1$
 in agreement with experimental ones \cite{tal}, \cite{mei}.
 Profiles with more than one extrema are also possible in the case
 of peaked $v_z$ profiles, localized  $v_\theta$ ones and $v_z
 v_{\theta}>0$  as demonstrated in Fig. \ref{fig:8b}.
 Experimental profiles of this kind were reported in Ref. \cite{mei} (Fig.
 (9) therein).
 \item The main contribution to $E_r$ comes from the velocity, to which is proportional,
 and particularly from the poloidal one (Fig. \ref{fig:5}).
 \item $E_r$ is sensitive to the relative orientation of
 $v_z$, $v_{\theta}$ and $B_z$; in particular, for $v_z v_\theta<0$   $|E_r|$ is larger than that
 for  $v_z v_\theta>0$. (Fig.  \ref{fig:5a}). Similar results hold
 for $E_r^\prime$ (Fig. \ref{fig:5b}) and $\omega_{{\bf E}\times{\bf B}}$ .
 \item  For extended velocity profiles with {\em $v_z\neq 0$}, an increase of $|s|$ results
 in an increase of $|E_r|$ (Fig. \ref{fig:6}), $|E_r^\prime|$ and $\omega_{{\bf E}\times{\bf B}}$.
  If $v_z=0$, however,  $|s|$ has no impact on $|E_r|$ and $|E_r^\prime|$, as
 can be seen by inspection of ${\bf E} = {\bf v}\times {\bf B} $,
 and very weak impact on
 $\omega_{{\bf E}\times{\bf B}}$. This result indicates that the presence of
 $v_z$  "activates" the impact of $s$ on $E_r$, $E_r^\prime$ and
 $\omega_{{\bf E}\times{\bf B}}$.
 \item An increase of the velocity shear nearly does not affect or even
 decreases the maximum $|E_r|$ (Fig. \ref{fig:7}) but increases
 $|E_r^\prime|$ (Fig. \ref{fig:8}).
 \end{itemize}


 \noindent
 4. {\large \em $\omega_{{\bf E}\times {\bf B}}$}
 \begin{itemize}
 \item  A typical profile of $\omega_{{\bf E}\times{\bf B}}$ has two large local maxima at the positions
 where the edges of the
 barrier are expected to be located in addition to other two smaller
 local ones (Fig. \ref{fig:9}). In most of the cases considered
 the maximum in the $s<0$ region is slightly larger than that in the $s>0$
 region.
(see  Fig. \ref{fig:9}).
In particular, for $B_z=\mbox{const.}$  at the point where $E_r'=0$
 one obtains:
 \beq
 \omega_{{\bf E}\times {\bf B}}=\Bigg|\frac{(1-s)\big(\epsilon
 \frac{\rho v_z}{q}-v_{\theta}\big)}{R_0q\Big[1+\big(\epsilon
 \frac{\rho}{q}\big)^2\Big]}\Bigg|
 \label{13}
 \eeq
 Eq. (\ref{13}) implies the following:
 \begin{enumerate}\item $\omega_{{\bf E}\times{\bf B}}$  depends on  the relative sign of
 $v_z$, $v_\theta$ and $B_z$, a result which we confirmed  by
 $\omega_{{\bf E}\times{\bf B}}$ profiles obtained via the symbolic computation programme.
 \item   The factor $(1-s)$ indicates that $\omega_{{\bf E}\times {\bf
 B}}$ for nearly shearless stellarator equilibria  may be lower than  that for tokamak equilibria
 with $s<0$.
 \item
 Despite the scaling $v_z\approx 10 v_\theta$,  for tokamak
 pertinent parametric values  the contributions of $v_z$-in connection with the term
 $\epsilon \rho v_z/q$- and $v_\theta$ to $\omega_{{\bf E}\times{\bf B}}$
 are of the same order of magnitude, a result indicating the importance of the
 poloidal velocity.
 \end{enumerate}

 \item  For extended velocities (large values of the parameter $h$ or/and peaked  $v_z$-
profile)  a percentage increase of $|s|$ in the barrier region results:
 \begin{enumerate}
 \item in approximately the same percentage increase of
 $\omega_{{\bf E}\times {\bf B}}$ if the velocity is purely axial (Fig. \ref{fig:10}).
 \item nearly does not affect the value of $\omega_{{\bf E}\times {\bf
 B}}$ if $v_{\theta}\neq 0$.
 \end{enumerate}

 %
 \item  An increase of 
 the  flow shear (variation of  the parameter $h$ from
0.1 to 0.001) causes a mean percentage increase of $\omega_{{\bf E}\times{\bf B}}$
 as large as 0.7 of that of the flow shear.  (Fig. \ref{fig:11}).
 \item The impact of a  variation of the $v_\theta$-shear on $\omega_{{\bf E}\times{\bf B}}$ is
 stronger than of the same variation of the $v_z$-shear.
 \item  The maximum increase of $\omega_{{\bf E}\times{\bf B}}$ is caused  by $v_\theta$
 in the case of non-vanishing peaked $v_z$ profiles.
 \item Inspection of  ${\bf v_{\bf E\times \bf B}}={\bf E}\times {\bf B}/B^2$ and (\ref{5})
 implies that $\omega_{{\bf E}\times{\bf B}}$ for a z-pinch is equal to that for an
 equilibrium with purely axial flow. The same equality is valid
 for a $\theta$-pinch in comparison with   an equilibrium with
 purely poloidal flow. In addition, it holds that
 \beq
 \omega_{{\bf E}\times{\bf B}-\mbox{z-pinch}}\approx 10\omega_{{\bf E}\times{\bf
 B}-\theta-\mbox{pinch}}.
 \label{st}
 \eeq
 \end{itemize}

\begin{center}
 {\large \bf 4. \ Conclusions}
\end{center}

 The self consistent study of cylindrical equilibria with reversed magnetic shear
 and sheared flow presented  in the previous sections led to the following conclusions:

 \begin{enumerate}
 \item For reversed magnetic shear profiles ($s<0$):
 \begin{itemize}
 \item The larger values of $|s|$ the  steeper the pressure
 profile.
 \item The axial current density profile become hollow.
 \item Strong reversed shear profiles formed  by appropriately large
 values of $\Delta q$ are associated with ''hole" axial
 current density profiles.
 \end{itemize}
These results are  consistent with experimental ones.
 \item Irrespective of the sign of $s$ the axial current
 density can reverse in the outer plasma region, the reversal point
 being shifted towards the plasma core as  $s$ takes lower
 negative values.
 \item  An increase  of either $|s|$ or the velocity  results generally in an increase of $|E_r|$, $|E_r^\prime|$ and
 $\omega_{{\bf E}\times{\bf B}}$.
\item  An increase  of the velocity shear results  in an increase of $|E_r^\prime|$ and
 $\omega_{{\bf E}\times{\bf B}}$.
\item For a given value of $|s|$, $\omega_{{\bf E}\times{\bf B}}$
 takes slightly larger values in the $s<0$ region than in the $s>0$
 region.
 \item $E_r$, $E_r^\prime$ and $\omega_{{\bf E}\times{\bf B}}$ are sensitive to the
 relative orientation of $v_\theta$, $v_z$ and $B_z$. In particular, they take larger values for
 $v_zv_{\theta}<0$ rather than for $v_zv_{\theta}>0$.
 %
 \item The presence of $v_z$ activates $s<0$, in the sense that for $v_z=0$,
 $E_r$ and $E_r'$
 are  $s$-independent. Also, for
 $v_z=0$, $s<0$ has very weak impact on $\omega_{{\bf E}\times {\bf
 B}}$.
 \item  The impact of the  poloidal flow and its shear on $E_r$, $E_r^\prime$  and $\omega_{{\bf E}\times{\bf B}}$ is
 stronger than that of the axial flow and the magnetic shear.
 \end{enumerate}

 {\em Presuming that $E_r$, $E_r^\prime$ and $\omega_{{\bf E}\times{\bf B}}$ are of relevance to the ITBs
 formation, the above results clearly indicate that
 the  reversed magnetic shear and the sheared flow  have  synergetic effects on 
 this formation with the flow, in particular the poloidal one, and its shear playing an
 important role.}

 \newpage
 \begin{center}
 {\large \bf  Figure captions}
 \end{center}

 \noindent
 Fig. \ref{fig:1}: \ SRS and WRS safety factor profiles associated with Eqs.
 (\ref{2}) and (\ref{3}), respectively.  It is noted that the finite slope of the
 WRS curve at $\rho=0$ may not  represent well the physical situation in the immediate vicinity of
the magnetic axis.
 \vspace{0.3cm}

\begin{figure}[!h]
 \begin{center}
 \psfrag{r}{$\rho$} \psfrag{p}{$\frac{P(\rho)}{P(0)}$}
 \psfrag{t}{$\Delta q=4 \ -$} \psfrag{d}{$\Delta q=14 \ \cdots$}
 \includegraphics[scale=1]{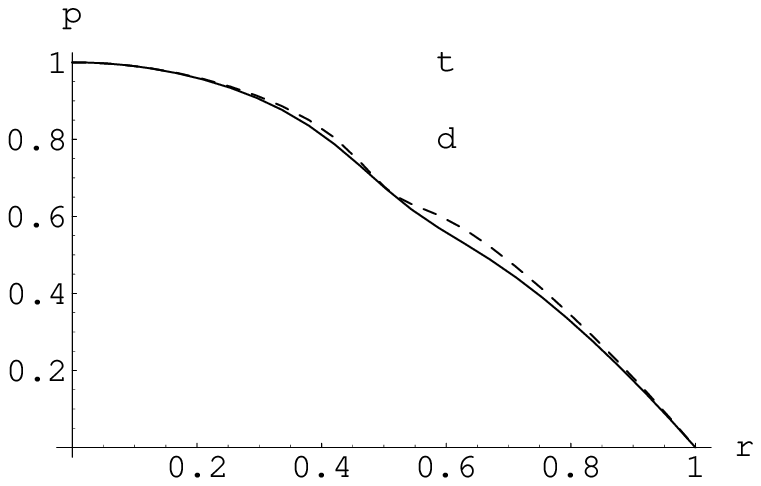}
 \caption{}
 \label{fig:2}
 \end{center}
 \end{figure}

 \noindent
 Fig. \ref{fig:2}: \ WRS pressure profiles  for $\Delta q =4$ and
 $\Delta q=14$.
 \vspace{0.3cm}

\begin{figure}[!h]
 \begin{center}
 \psfrag{j}{$\frac{J_z(\rho)}{J_z(0)}$}
 \psfrag{r}{$\rho$}
 \psfrag{w}{WRS $-$}\psfrag{s}{SRS $\cdots$}
 \includegraphics[scale=1]{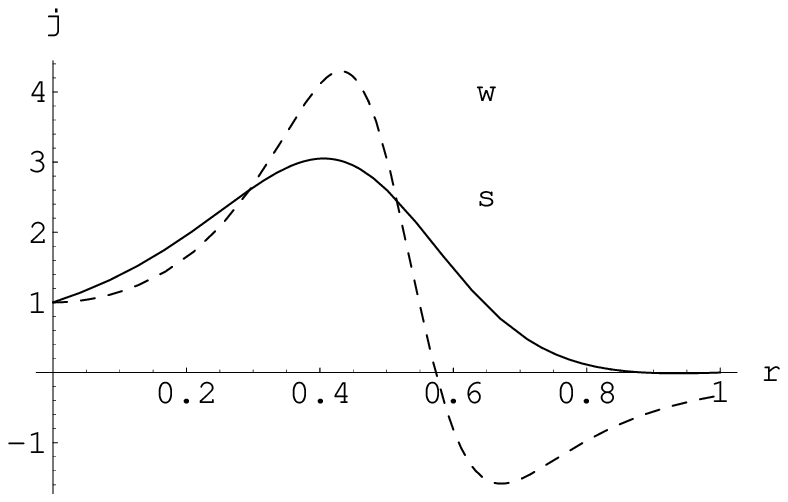}
 \caption{}
 \label{fig:3}
 \end{center}
 \end{figure}

 \noindent
 Fig. \ref{fig:3}: \ Toroidal current density profiles for $\Delta q=4$.
   It is noted that the finite slope of the
 WRS curve at $\rho=0$ may not  represent well the physical situation in the immediate vicinity of
the magnetic axis.
 \vspace{0.3cm}

\begin{figure}[!h]
 \begin{center}
 \psfrag{j}{$\frac{J_z(\rho)}{J_z(0)}$}
 \psfrag{r}{$\rho$}
 \includegraphics[scale=1]{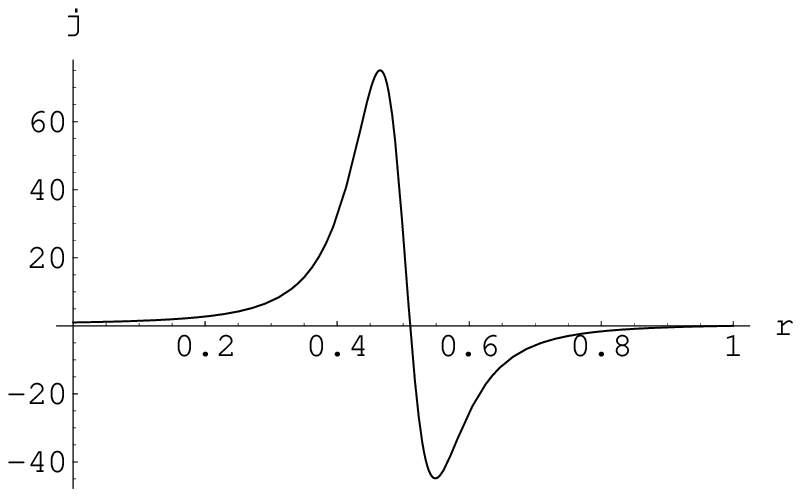}
 \caption{}
 \label{fig:4}
 \end{center}
 \end{figure}

\begin{figure}[!h]
 \begin{center}
 \psfrag{e}{$\frac{E_r(\rho)}{E_r(0.5)}$}
 \psfrag{r}{$\rho$}
 \includegraphics[scale=1]{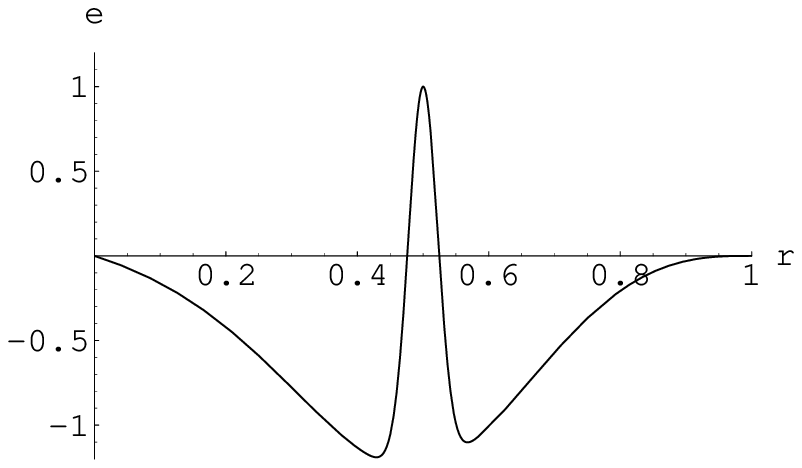}
 \caption{}
 \label{fig:8b}
 \end{center}
 \end{figure}

\begin{figure}[!h]
 \begin{center}
 \psfrag{e}{$|E_{r-norm}(\rho)|$}
 \psfrag{r}{$\rho$}
 \psfrag{z}{$v_{\theta}=0 \ -$}\psfrag{t}{$v_z=0 \ \cdots$}
 \includegraphics[scale=1]{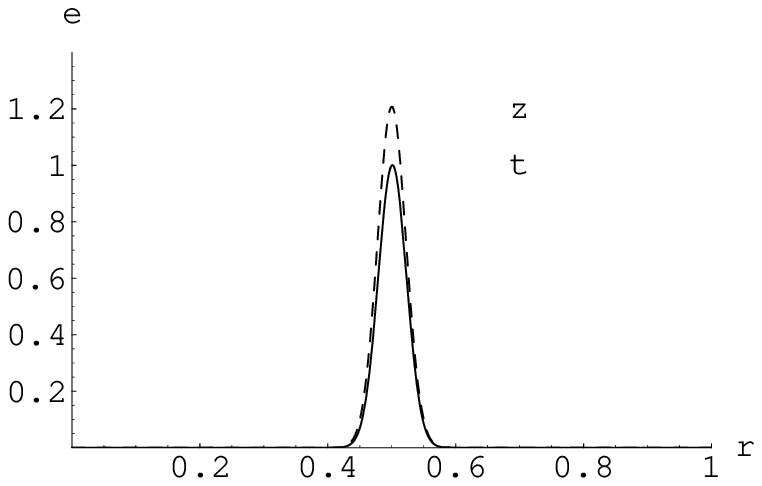}
 \caption{}
 \label{fig:5}
 \end{center}
 \end{figure}

\begin{figure}[!h]
 \begin{center}
 \psfrag{e}{$|E_{r-norm}(\rho)|$}
 \psfrag{r}{$\rho$}
 \psfrag{g}{$v_{\theta}\cdot v_z>0 \ -$}\psfrag{l}{$v_{\theta}\cdot v_z<0 \ \cdots$}
 \includegraphics[scale=1]{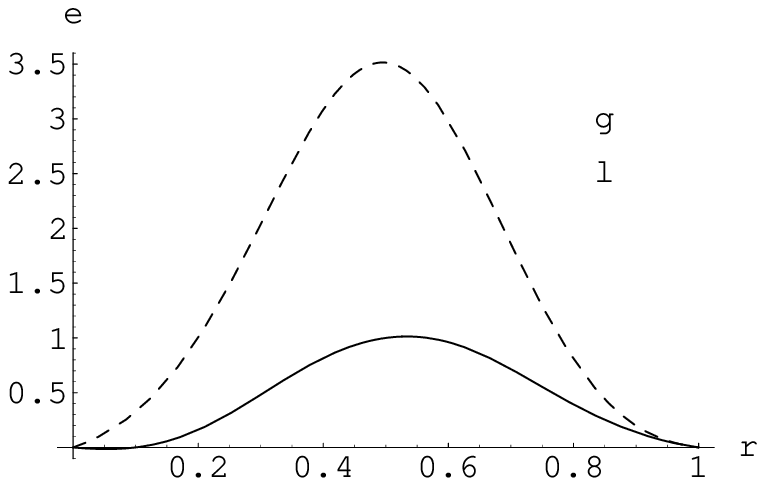}
 \caption{}
 \label{fig:5a}
 \end{center}
 \end{figure}

\begin{figure}[!h]
 \begin{center}
 \psfrag{e}{$E_{r-norm}^\prime(\rho)$}
 \psfrag{r}{$\rho$}
 \psfrag{g}{$v_{\theta}\cdot v_z>0 \ -$}\psfrag{l}{$v_{\theta}\cdot v_z<0 \ \cdots$}
 \includegraphics[scale=1]{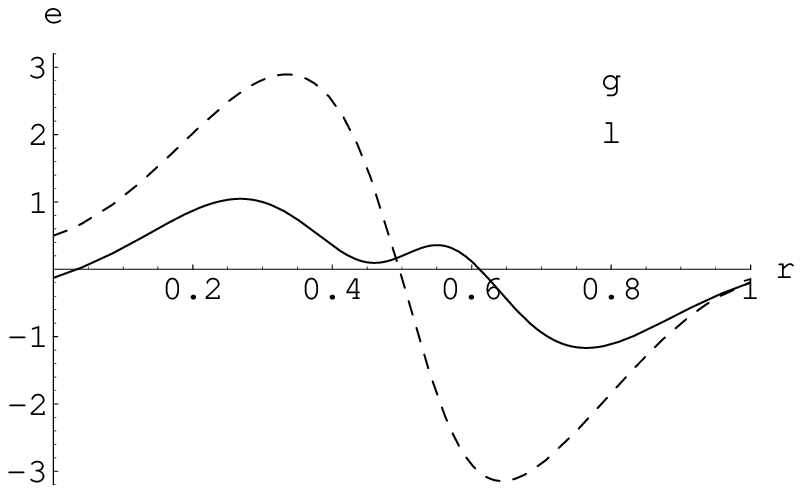}
 \caption{}
 \label{fig:5b}
 \end{center}
 \end{figure}

\begin{figure}[!h]
 \begin{center}
 \psfrag{e}{$|E_{r-norm}(\rho)|$}
 \psfrag{r}{$\rho$}
 \psfrag{t}{$\Delta q=4 \ -$}
 \psfrag{d}{$\Delta q=14 \ \cdots$}
 \includegraphics[scale=1]{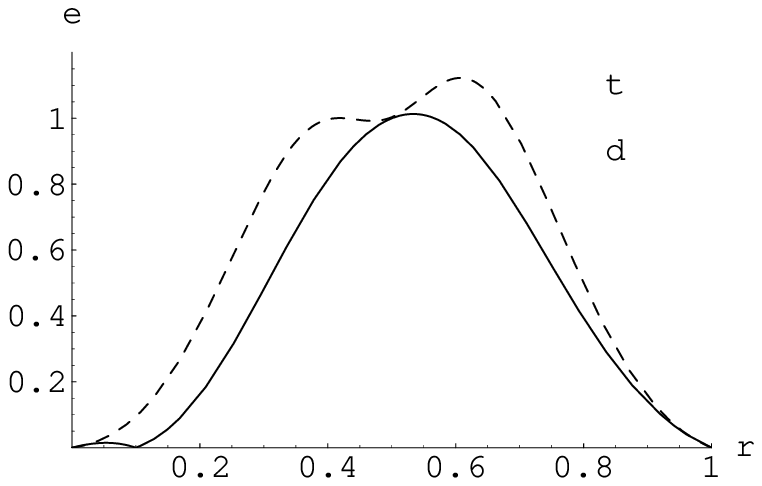}
 \caption{}
 \label{fig:6}
 \end{center}
 \end{figure}

\begin{figure}[!h]
 \begin{center}
 \psfrag{e}{$|E_{r-norm}(\rho)|$}
 \psfrag{r}{$\rho$}
 \psfrag{x}{$h=0.1 \ -$}
 \psfrag{l}{$h=0.001 \ \cdots$}
 \includegraphics[scale=1]{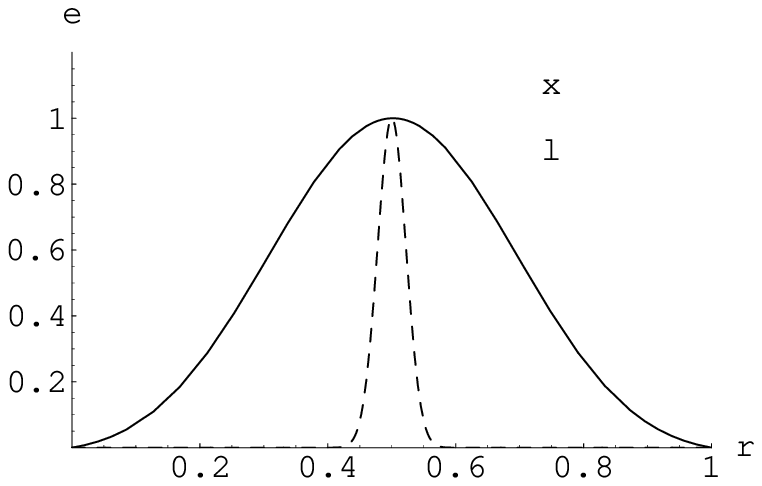}
 \caption{}
 \label{fig:7}
 \end{center}
 \end{figure}

\begin{figure}[!h]
 \begin{center}
 \psfrag{se}{$E_{r-norm}^\prime(\rho)$}
 \psfrag{r}{$\rho$}
 \psfrag{x}{$h=0.1 \ -$}
 \psfrag{l}{$h=0.001 \ \cdots$}
 \includegraphics[scale=1]{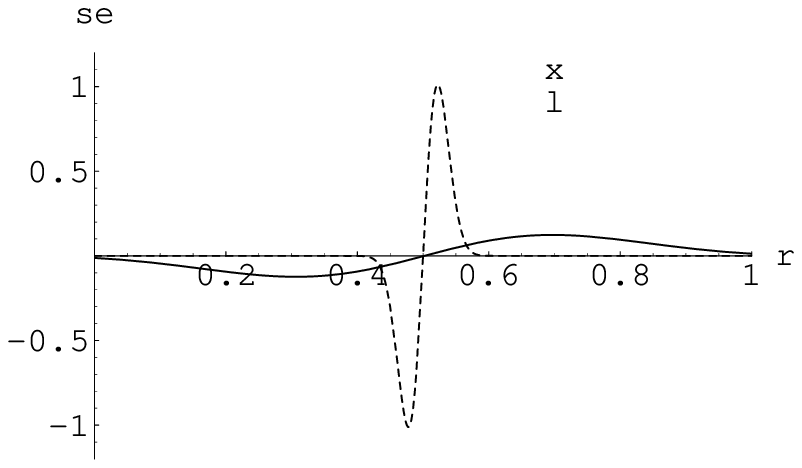}
 \caption{}
 \label{fig:8}
 \end{center}
 \end{figure}

\begin{figure}[!h]
 \begin{center}
 \psfrag{o}{$\frac{\omega_{{\bf E}\times {\bf B}}(\rho)}{\omega_{{\bf E}\times {\bf B}}(0.45)}$}
 \psfrag{r}{$\rho$}
 \includegraphics[scale=1]{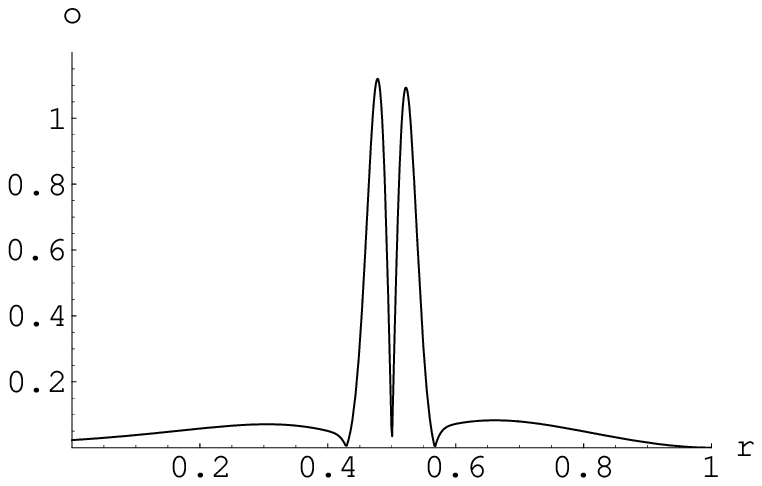}
 \caption{}
 \label{fig:9}
 \end{center}
 \end{figure}

\begin{figure}[!h]
 \begin{center}
 \psfrag{o}{$\omega_{{\bf E}\times {\bf B} -norm}(\rho)$}
 \psfrag{r}{$\rho$}
 \psfrag{t}{$\Delta q=4 \ -$}
 \psfrag{d}{$\Delta q=14 \ \cdots$}
 \includegraphics[scale=1]{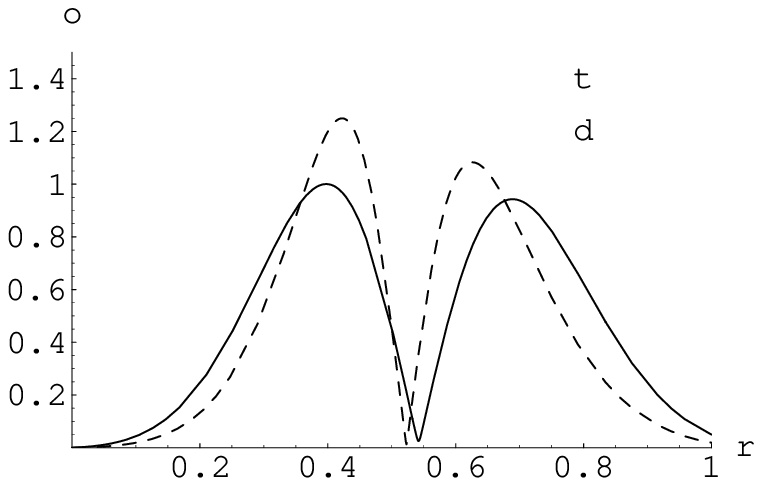}
 \caption{}
 \label{fig:10}
 \end{center}
 \end{figure}

\begin{figure}[!h]
 \begin{center}
 \psfrag{o}{$\omega_{{\bf E}\times {\bf B} -norm}(\rho)$}
 \psfrag{r}{$\rho$}
 \psfrag{x}{$h=0.1 \ -$}
 \psfrag{l}{$h=0.001 \ \cdots$}
 \includegraphics[scale=1]{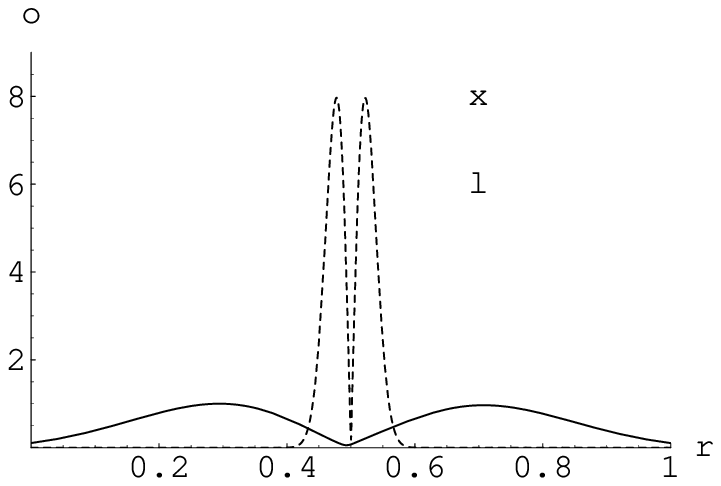}
 \caption{}
 \label{fig:11}
 \end{center}
 \end{figure}

 \noindent
 Fig. \ref{fig:4}: \ Toroidal current density profile for WRS, $q_c=102$ and $\Delta q=100$ that demonstrates
 the current "hole' in the core region.
 \vspace{0.3cm}

 \noindent
 Fig. \ref{fig:8b}: \ Electric field profile for WRS with $v_z$ peaked and $v_{\theta}$ localized
 having three local extrema.
 \vspace{0.3cm}

 \noindent
 Fig. \ref{fig:5}: \ Two  $E_r$-profiles the one with $v_z=0$ and the other with $v_{\theta}=0$
 for SRS and Gaussian-like velocity profiles.
 $E_r$ is
 normalized with respect to its value at $\rho =0.5$ for $v_z=0$ .
 \vspace{0.3cm}

 \noindent
 Fig. \ref{fig:5a}: \ Two $E_r$-profiles    for $v_z$ peaked and SRS, the one with
 $v_\theta \cdot v_z>0$ and the other with $v_\theta \cdot v_z<0$.
 The profiles are normalized with respect to the first case at
 $\rho =0.5$.
 \vspace{0.3cm}

 \noindent
 Fig. \ref{fig:5b}: \ Two profiles of $E_r'$ with $v_z$ peaked and  SRS, the
 one with $v_\theta \cdot v_z>0$ and the other with $v_\theta \cdot v_z<0$.
 The profiles are normalized with respect to the first case at
 $\rho =0.3$.
 \vspace{0.3cm}

 \noindent
 Fig. \ref{fig:6}: \ Profiles of $E_r$ with peaked axial and extended poloidal
 velocities for WRS and  two different values of $\Delta q$. The profiles are
 normalized with respect to the case with $\Delta q=4$ at
 $\rho=0.5$.
 \vspace{0.3cm}

 \noindent
 Fig. \ref{fig:7}: \ Profiles of $E_r$ with $v_z=0$ for SRS and either extended ($h=0.1$)
 or localized ($h=0.001$) poloidal velocity. The profiles are
 normalized with respect to the first case at $\rho =0.5$.
 \vspace{0.3cm}

 \noindent
 Fig. \ref{fig:8}: \ Two profiles of $E_r'$ for SRS  with $v_z=0$ the one for extended  ($h=0.1$)
 and the other for localized  ($h=0.001$) poloidal velocities. The profiles are
 normalized with respect to the second case at $\rho =0.55$.
 \vspace{0.3cm}

 \noindent
 Fig. \ref{fig:9}: \ Typical $\omega_{{\bf E}\times {\bf B}}$-profile  for WRS, peaked
 axial
 and localized poloidal velocities.
 \vspace{0.3cm}

 \noindent
 Fig. \ref{fig:10}: \ Profiles of  $\omega_{{\bf E}\times {\bf B}}$ for WRS, peaked
 axial velocity, and either $\Delta q=4$ or $\Delta q=14$. The profiles are
 normalized with respect to the first case at $\rho =0.45$.
 \vspace{0.3cm}

 \noindent
 Fig. \ref{fig:11}: \ $\omega_{{\bf E}\times {\bf B}}$-profile for SRS,
 Gaussian-like  axial
 and poloidal velocity components   both either
 extended ($h=0.1$) or localized ($h=0.001$). The profiles are
 normalized with respect to the first case at $\rho =0.3$.

 \newpage

 \end{document}